\RequirePackage{fix-cm,amsmath}
\documentclass{svjour3}                     
\smartqed  
\usepackage{color,graphicx,amsmath,amsfonts}
\begin{document}

\title{Linear response of one-dimensional liquid $^4$He to external perturbations}

\titlerunning{Density response of 1D $^4$He}

\author{M.~Motta    \and
        G.~Bertaina \and
        E.~Vitali   \and
        D.E.~Galli  \and
        M.~Rossi}

\authorrunning{Motta et al.}

\institute{M. Motta and E. Vitali \at  
           Department of Physics, The College of William and Mary, Williamsburg, Virginia 23187, USA\\
           \email{mmotta@wm.edu}  
           \and
           D.E. Galli and G. Bertaina \at
           Dipartimento di Fisica, Universit\`a degli Studi di Milano, via Celoria 16, I-20133 Milano, Italy\\
           \email{gianluca.bertaina@unimi.it} 
           \and
           M. Rossi \at
           Scuola Normale Superiore, Piazza dei Cavalieri 7, I-56126 Pisa, Italy;\\
           International Center for Theoretical Physics (ICTP), Strada Costiera 11, I-34154 Trieste, Italy    
}

\date{Received: date / Accepted: date}

\maketitle

\begin{abstract}
We study the response of one-dimensional liquid $^4$He to weak perturbations relying on the dynamical
structure factor, $S(q,\omega)$, recently obtained via {\it ab-initio} techniques 
[{\it Phys. Rev. Lett.} {\bf 116}, 135302 (2016)].
We evaluate the drag force, $F_v$, experienced by an impurity moving along the system with velocity 
$v$ and the static response function, $\chi(q)$, describing the density modulations induced by a 
periodic perturbation with wave vector $q$.
 

\end{abstract}

\section{Introduction}
\label{intro}

One-dimensional (1D) systems occupy a unique place in the realm of many-body quantum mechanics,
due in particular to the loosened distinction between the behavior of fermions and bosons with 
hard-core repulsive interaction.
Electronic transport properties in quasi-1D have been extensively investigated in various systems 
such as fractional quantum Hall edge states \cite{chang_1996}, carbon nanotubes 
\cite{yao_1999,bockrath_1999}, conducting conjugated polymer nanowires \cite{aleshin_2004}, and 
semiconductors nanowires \cite{zotov_2000}.
Also quasi-1D bosonic systems realized with ultracold atoms confined in cigar-shaped traps have 
attracted considerable attention \cite{cazalilla_2011}. 
1D behavior has also been theoretically predicted and experimentally observed
to play a crucial role for the characterization of $^4$He atoms moving inside 
dislocation lines in crystalline helium samples \cite{boninsegni_2007,vekhov_2012,vekhov_2014}, or
confined inside nanopores \cite{wada_2001,wada_2008,yamashita_2009,delmaestro_2010,delmaestro_2011,matsushita_2015}.

The low-energy properties of a vast class of 1D systems, including liquid $^4$He, are well 
captured by the Tomonaga-Luttinger liquid theory (TLL) \cite{tomonaga_1950,luttinger_1963,haldane_1981}.
The TLL theory is an effective low-energy field theory governed by an exactly solvable quadratic 
Hamiltonian that depends only on two parameters, namely the sound velocity $c$ and Luttinger 
parameter $K_L$, which characterize the large-distance behavior of the correlation functions and the
thermodynamic properties of the system. 
For spinless Galilean-invariant systems, as the ones we are going to consider here, these two 
parameters are related via $c = v_F/K_L$, where $v_F=\hbar \pi \rho/m$ is the Fermi velocity of a 1D
ideal Fermi gas with the same density $\rho= L/N$ of the system.
Thus the properties of the system depends only on a single parameter, $K_L$, that turns out to be
proportional to the square root of the compressibility $\kappa_S$: $K_L=\sqrt{\hbar^2\pi^2\rho^3\kappa_S/m}$.

Since the Mermin-Wagner theorem \cite{mermin_1966} prevents the existence of Bose-Einstein condensation
in 1D systems with short-range velocity-independent interaction, the standard picture of superfluidity
relying on the order parameter provided by the condensate wavefunction \cite{leggett_1999} has to
be generalized \cite{eggel_2011}.
However, in the TLL phase, correlation functions feature a power-law decay \cite{haldane_1981}
and the system is superfluid in the sense that it displays a quasi-off-diagonal long-range order
\cite{cazalilla_2011,eggel_2011}.

In this paper we characterize the dynamical properties of 1D liquid $^4$He by computing the friction 
or drag force, $F_v$, exerted on a particle of mass $m$ moving with velocity $v$ inside the system, 
which provides a generalization of Landau's celebrated criterion of superfluidity 
\cite{landau_1941,landau_1947}.
We determine also the static density response function, $\chi(q)$, which is of paramount importance 
for the response of the system to a periodic potential, relevant for the interpretation of experiments 
\cite{boninsegni_2007,taniguchi_2010,taniguchi_2011} and for density functionals theories 
\cite{dalfovo_1995,minoguchi_2011,minoguchi_2012,minoguchi_2013}.

Both $F_v$ and $\chi(q)$ are obtainable from the knowledge of the dynamical structure factor,
$S(q,\omega)$, that we have recently achieved via an {\it ab-initio} full microscopic approach 
\cite{bertaina_2016}, by combining the Genetic Inversion via Falsification of Theories (GIFT) algorithm 
\cite{vitali_2010,rossi_2012,nava_2013,nava_2013b,arrigoni_2013,rota_2013,rota_2014,molinelli_2016} 
to perform the analytic continuation of imaginary-time correlation functions computed via the exact 
zero-temperature Path Integral Ground State (PIGS) method \cite{sarsa_2000,galli_2003,rossi_2009}.

\section{Drag Force}
\label{sec:drag}
 
According to Landau \cite{landau_1941,landau_1947}, an obstacle in a fluid, moving with velocity $v$, 
may cause transitions from the ground state to excited states lying on the line 
$\epsilon(k) = \hbar v \, k$ in the energy-momentum space. 
If all the spectrum is above this line, the motion cannot excite the system, and the flow of the
impurity is frictionless. 
However, even when the line intersects the spectrum, the transition probabilities to these states 
can be strongly suppressed due to the interaction or to the external perturbing potential. 
In this case, the drag force provides a quantitative measure of superfluidity.

The drag force, $F_v$, exerted on an impurity moving with velocity $v$ in a 1D medium, can be computed
relying on Fermi's golden rule, by expressing the energy loss per unit time due to the impurity, 
$\frac{d \mathcal{E}}{dt}=-F_v v$, as \cite{cherny_2011}:
\begin{equation}
 \label{eq:Eloss}
  \frac{d \mathcal{E}}{dt} = -\sum_{q (= k_f - k_i)} p(k_i \to k_f) \, \hbar\omega_q 
                           = -\int dq \, |V_q|^2 \, \rho \, \hbar\omega_q \, S(q,\omega_q)
\end{equation}
where $p(k_i \to k_f)$ is the rate for the scattering process, 
$\omega_q = \hbar(k_f^2 - k_i^2)/2m = qv - \hbar q^2/2m$ 
and $V_{q}$ the Fourier transform of the impurity-medium interaction $V$.
For a heavy impurity, $v \gg \hbar q/m$, \eqref{eq:Eloss} simplifies to:
\begin{equation}
 \label{eq:dragformula}
  \frac{d \mathcal{E}}{dt} = -\left( \int_0^{\infty} dq |V_q|^2 \rho \, \hbar q \, S(q,qv) \right) v 
                           = -F_v v
\end{equation}
Equation \eqref{eq:dragformula} shows 
that knowledge of $S(q,\omega)$ and of $V_q$ grants the ability of computing the drag force. 
The relationship between the concept of drag force and Landau's criterion for superfluidity is 
readily understood: if $S(q,\omega)$ is concentrated above the straight line $\hbar v_0 q$ in the 
momentum-energy plane, then $F_v \equiv 0$ for all velocities $v < v_0$.
On the other hand, the integral can be very small or vanish even if the spectrum lies below the line
$\hbar v_0 q$, but the excitation probabilities are suppressed \cite{cherny_2011}, for instance 
because $V_q$ takes non-zero values only in a finite region of momentum space.
In a broad class of 1D systems, like the ideal Fermi gas, $^4$He at high density \cite{bertaina_2016} 
and hard rods \cite{motta_2016}, $S(q,\omega)$ touches the $\omega = 0$ line with finite weight at 
$2 k_F$ ($k_F = \pi\rho$ being the Fermi wave-vector), whence the possibility of a dissipationless 
flow is by no means obvious.

\begin{figure}[t]
 \centering
  \includegraphics[width=0.7\textwidth]{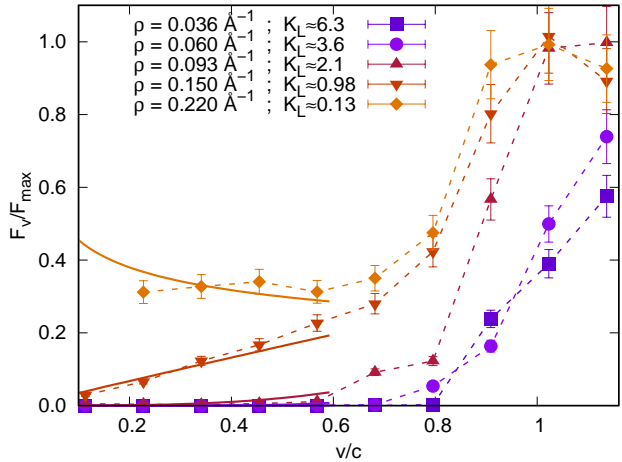}
  \caption{Drag force $F_v$ as a function of the ratio $v/c$, compared to the Luttinger liquid
           theory prediction for small $v$ (solid lines) and normalized to the maximum value $F_{max}$. 
           Corresponding values of the Luttinger parameter are also indicated. Errorbars are 
           estimated from the typical variance of the reconstructed GIFT spectra.
           (Color figure online).}
\label{fig:drag}
\end{figure}

To compute the drag force (\ref{eq:dragformula}), we have approximated $V_q$ by using a contact 
potential with the same scattering length of the repulsive part of the $^4$He interaction (as in 
Ref. \cite{kalos_1974}), and the $S(q,\omega)$ estimated in Ref. \cite{bertaina_2016}.
Our results are shown in Fig. \ref{fig:drag}.
We observe an increase of $F_v$ with the velocity, and a saturation to a maximum value 
$F_{max}$ at $v \simeq c$.
The increase is consistent with Luttinger liquid theory \cite{cherny_2011,astrakharchik_2004}. 
In fact, for a slow impurity $v \ll c$, the most important contribution to \eqref{eq:dragformula} 
comes from the region $q \simeq 2 k_F$. 
As shown in Ref.\cite{astrakharchik_2004}, for  $q \to 2 k_F$ and $\omega \to 0$, $S(q,\omega)$ has 
the power-law behavior:
\begin{equation}
 \label{eq:powerlawLUT}
 S(q,\omega) \propto \omega^{2 (K_L-1)}( 1 - x^2)^{K_L-1}
\end{equation}
being $x = c(q-2k_F)/\omega$, with the constraint $|x| \le 1$.  
Inserting \eqref{eq:powerlawLUT} into \eqref{eq:dragformula} yields the following expression for the 
drag force:
\begin{equation}
 F_v \propto \int_{-1}^{\min\left( \frac{c}{v},1 \right)} \frac{dx}{\left(1 - \frac{v}{c} x \right)^3} 
             \left( \frac{\frac{v}{c}}{1 - \frac{v}{c} x} \right)^{2(K_L-1)}(1-x^2)^{K_L-1}.
\end{equation}
For $v \ll c$, the approximation $1 - \frac{v}{c} x \simeq 1$ is accurate and one finds 
$F_v = F_0 (v/c)^{2K_L-1}$ \cite{cherny_2011,astrakharchik_2004}, that compares satisfactory with 
our estimates of $F_v$.

We remark that the drag force is non-vanishing at any $v$, but for $K_L > 1/2$ its power-law behavior 
determines a superfluid response of the system, allowing for impurities to flow with small dissipation \cite{giamarchi_1988}. 
Notice also that, for $K_L<1/2$ (e.g. at the density $\rho=0.220$\AA$^{-1}$),
the previous analytical expression implies a divergence of $F_v$ for small velocities; nonetheless, the physical dissipated power in Eq.\eqref{eq:dragformula} 
is always vanishing at small velocity, since it behaves as $\dot{E} \propto v^{2K_L}$.
Our estimate of $F_v$ is perturbative, therefore it is relevant for soft impurities such as small 
geometry deformations in quasi-1D systems.
It is clear that for hard-core impurities (like $^3$He pinned to $^4$He dislocations 
\cite{galli_2008,chan_2013}) the superfluid response is completely suppressed.

\section{Static density response function}
\label{sec:chiq}

\begin{figure}[t]
 \centering
  \includegraphics[width=0.7\textwidth]{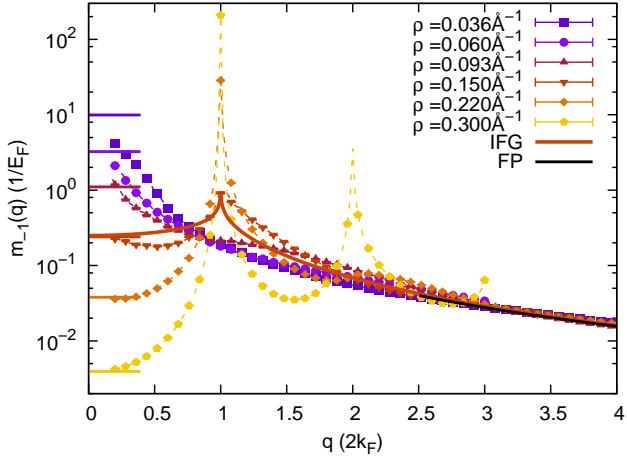}
  \caption{First negative momentum $m_{-1}(q)$ in units of the inverse Fermi energy $1/E_F$, as 
  		   function of $q$ in units of $2k_F$.
           Small horizontal lines at $q \to 0$ correspond to the asymptotic estimate based on the 
           Feynman approximation.
           The static response function for the ideal Fermi gas (IFG) and the free particle limit (FP)
           are also reported as solid lines.
           (Color figure online).}
 \label{fig:chiq}
\end{figure}

The static density response function characterizes the effect of a static periodic perturbation on a 
homogeneous system \cite{sugiyama_1992,bowen_1994,moroni_1995,depalo_2003}, and can be computed 
relying on the Hellman-Feynman theorem \cite{moroni_1995} or from the first negative moment, $m_{-1}(q)$, 
of $S(q,\omega)$:
\begin{equation}
 \chi(q) = - \frac{2\rho}{\hbar} \int_{0}^{\infty} d\omega \, \frac{S(q,\omega)}{\omega} = 
           - \frac{2\rho}{\hbar} m_{-1}(q).
\end{equation}
Our results are shown in Fig. \ref{fig:chiq}.  

In the low-momentum regime, according to the TLL theory the dynamical structure factor 
is concentrated around the dispersion relation $\omega_{LL}(q) = c |q|$, 
and the Feynman approximation $S(q,\omega) = S(q) \, \delta(\omega-\omega_{LL}(q))$ can be safely 
assumed, with $S(q) \simeq K_L \, \frac{q}{2 k_F}$.
Therefore:
\begin{equation}
 m_{-1}(q) = \frac{1}{E_F} \, \frac{S(q)^2}{\left( \frac{q}{k_F} \right)^2}
          \simeq \frac{1}{E_F} \, \frac{K_L^2}{4}
\end{equation}
being $E_F = \hbar^2k_F^2/2m$ the Fermi energy.
This relation is equivalent to the well known compressibility sum rule obeyed by $\chi(q)$ in higher
dimensions \cite{caupin_2008}.
The resulting values for the different considered densities are reported as solid lines in 
Fig. \ref{fig:chiq}, instead of the obtained $\chi(q)$, that at low momenta is plagued by
frequency-discretization effects unavoidable with the actual implementation of GIFT.

As clearly visible in Fig. \ref{fig:chiq}, $\chi(q)$ features peaks at momenta which are integer 
multiples of $2k_F$,
at the densities $\rho = 0.22$~\AA$^{-1}$ and $\rho = 0.30$~\AA$^{-1}$, where $K_L$ is found to be
$K_L = 0.389(2)$ and $K_L = 0.1255(5)$ respectively \cite{bertaina_2016}.
The presence of such peaks can be easily justified in the light of the TLL theory, with the 
following heuristic argument. 
Assuming that $S(q,\omega)$ has the low-frequency power-law divergence 
$S(q=2jK_F,\omega) \propto \omega^{2(j^2 k_L-1)}$ at $q=2jk_F$ 
\cite{bertaina_2016,astrakharchik_2004,luther_1974,castro_1994}, 
and has support above the energy $\hbar\omega_{sc,j}= 4 E_F j^2/N$ of the supercurrent state obtained 
applying a boost of momentum $2jk_F$ to the ground state \cite{cherny_2011}, one can easily prove 
\cite{mazzanti_2008} the presence of peaks in $S(q=2jk_F)$: 
\begin{equation}
 S(q=2jk_F) = \int_{\omega_{sc,j}}^\infty d\omega \, S(q,\omega) \propto 
                    \omega_{sc,j}^{2(j^2 k_L-1)+1} \propto N^{1-2 j^2 k_L}
\end{equation}
because $N^{1-2 j^2 k_L}$ diverges for $K_L < \frac{1}{2j^2}$ in the thermodynamic limit. 
Applying the same argument to $m_{-1}(q)$ shows that:
\begin{equation}
\label{eq:chiqdiverge}
  m_{-1}(q=2jk_F) = \int_{\omega_{sc,j}}^\infty d\omega \, \frac{S(q,\omega)}{\omega} \propto 
                    \omega_{sc,j}^{2(j^2 k_L-1)} \propto N^{2-2 j^2 k_L}
\end{equation}
diverging in the thermodynamic limit provided that $K_L < \frac{1}{j^2}$.
Equation \eqref{eq:chiqdiverge} explains the presence of peaks in $\chi(2k_F)$ at the density 
$\rho = 0.220$~\AA$^{-1}$, and in $\chi(2 k_F), \chi(4k_F)$ at $\rho = 0.300$~\AA$^{-1}$.

The $S(q,\omega)$ of 1D Bose systems with repulsive interaction manifests a pseudo particle-hole,
typical of Fermi systems, due to a peculiar interplay between dimensionality and repulsion 
\cite{bertaina_2016}.
This suggests that also $\chi(q)$ should display a behavior similar to the static density response
function of the ideal Fermi gas, known also as static Lindhard function \cite{vignale_2005}, that we 
report in Fig. \ref{fig:chiq} as a solid line.
In the large $q$ limit, our results collapse on the asymptotic behavior $m_{-1}(q) \simeq \frac{k_F^2}{q^2}$ 
as expected also in higher dimensions (free particle limit) \cite{caupin_2008}.

\section{Conclusions}
\label{sec:concl}

We addressed the calculation of linear response functions of one-dimensional $^4$He, relying on 
state-of-art Quantum Monte Carlo calculations of the dynamical structure factor of the system. 
In particular, we estimated the force experienced by an impurity moving through the sample and weakly
interacting with the helium atoms. 
This calculation is very interesting since it addresses the fate of the celebrated Landau criterion
for superfluidity in a scenario in which the Mermin-Wagner theorem rules out the possibility of Bose
Einstein condensation.
Our results display a power-law behavior of the drag force as a function of the velocity of the 
impurity, showing thus a quasi-superfluid response of the system, consistently with the Luttinger 
liquid paradigm.

We also computed the static density response function $\chi(q)$ of the system, providing the linear
response of the helium atoms to an external periodic potential.
For interacting quantum systems, this property is notoriously hard to access, requiring either the 
estimation of dynamical properties or the introduction of an external potential \cite{motta_2015}.
We suggest interpretation of our data relying on known properties of the dynamical structure factor 
$S(q,\omega)$, such as the compressibility sum rule in the $q\to0$ limit and the free-particle 
$q\to\infty$ limit.
Varying the density, we observe a crossover from a dilute regime with a smooth static density 
response function to a quasi-solid high-density regime, where $\chi(q)$ displays peaks whose height 
increases with the size of the system.

Although our results are valid for a strictly 1D configuration, while experiments with $^4$He atoms deal with quasi-1D geometries, 
such as nanopores or dislocation lines, we hope our findings can provide useful insight and a firm limiting case for realistic situations. 
In particular, helium in nanopores has been observed to manifest anomalous heat capacity \cite{wada_2001} and superfluid response in torsional oscillator experiments  \cite{ikegami_2007}, with respect to the 2D behavior which is expected from the adsorbed films on the nanopore walls. Extinction of superfluid effects in the limiting 1D case (very small nanopore radius) is elucidated in the present work, depending on the linear density.

\begin{acknowledgements}
We acknowledge the CINECA and the Regione Lombardia award LI03p-UltraQMC, under the LISA 
initiative, for the availability of high-performance computing resources and support.
M.M. and E.V. acknowledge support from the Simons Foundation and NSF (Grant no. DMR-1409510).
M.R. acknowledges the EU Horizon 2020 FET QUIC project for fundings. G.B. acknowledges support from the
University of Milan through Grant No. 620, Linea 2 (2015).
\end{acknowledgements}


\begin{thebibliography}{unsrt}

\bibitem{chang_1996}		A.M. Chang, L.N. Pfeiffer and K.W. West,
							{\em Phys. Rev. Lett.} {\bf 77}, 2538 (1996).

\bibitem{yao_1999}			Z. Yao, H.W.C. Postma, L. Balents and C. Dekker,
							{\em Nature} {\bf 402}, 273 (1999).
							
\bibitem{bockrath_1999}		M. Bockrath, D.H. Cobden, J. Lu, A.G. Rinzler, R.E. Smalley, L. Balents and P.L. McEuen,
							{\em Nature} {\bf 397}, 598 (1999).

\bibitem{aleshin_2004}		A.N. Aleshin, H.J. Lee, Y.W. Park and K. Akagi
							{\em Phys. Rev. Lett.} {\bf 93}, 196601 (2000).

\bibitem{zotov_2000}		S.V. Zaitsev-Zotov, Y.A. Kumzerov, Y.A. Firsov and P. Monceau,
							{\em Journal of Physics: Condensed Matter} {\bf 12}, L303 (2000).

\bibitem{cazalilla_2011} 	M.A. Cazalilla, R. Citro, T. Giamarchi, E. Orignac and M. Rigol,
							{\em Rev. Mod. Phys.} {\bf 83}, 1405 (2011).

\bibitem{boninsegni_2007}	M. Boninsegni, A.B.	Kuklov, L. Pollet, N.V. Prokof'ev, B.V. Svistunov and M. Troyer, 
							{\em Phys. Rev. Lett.} {\bf 99}, 035301 (2007).

\bibitem{vekhov_2012}		Y. Vekhov and R.B. Hallock,
							{\em Phys. Rev. Lett.} {\bf 109}, 045303 (2012).

\bibitem{vekhov_2014}		Y. Vekhov and R.B. Hallock,
							{\em Phys. Rev. B} {\bf 90}, 134511 (2014).

\bibitem{wada_2001}		N. Wada, J. Taniguchi, H. Ikegami, S. Inagaki, and Y. Fukushima,
							{\em Phys. Rev. Lett.} {\bf 86}, 4322 (2001).

\bibitem{wada_2008}		N. Wada and M. W. Cole,
							{em J. Phys. Soc. Jpn.} {\bf 77}, 111012 (2008).
							
\bibitem{yamashita_2009}	K. Yamashita and D.S. Hirashima,
							{\em Phys. Rev. B} {\bf 79}, 14501 (2009).

\bibitem{delmaestro_2010}	A. Del Maestro and I. Affleck,
							{\em Phys. Rev. B} {\bf 82}, 060515 (2010).

\bibitem{delmaestro_2011}	A. Del Maestro, M. Boninsegni and I. Affleck,
							{\em Phys. Rev. Lett.} {\bf 106}, 105303 (2011).


\bibitem{matsushita_2015}	T. Matsushita, A. Shinohara, M. Hieda, and N. Wada,
							{\em J. Low. Temp. Phys.} {\bf 183}, 273 (2015).

							
\bibitem{tomonaga_1950} 	S.-I. Tomonaga,
							{\em Prog. Theor. Phys.} {\bf 5}, 544 (1950).
				
\bibitem{luttinger_1963} 	J.M. Luttinger,
							{\em J. Math. Phys.} {\bf 4}, 1154 (1963).
							
\bibitem{haldane_1981} 		F.D.M. Haldane,
							{\em Phys. Rev. Lett.} {\bf 47}, 1840 (1981).

\bibitem{mermin_1966}		N.D. Mermin and H.	Wagner,
						    {\em Phys. Rev. Lett.}, {\bf 17}, 1133 (1966).

\bibitem{leggett_1999}		A.J. Leggett,
							{\em Rev. Mod. Phys.} {\bf 71}, 318 (1999).
							
\bibitem{eggel_2011}	 	T. Eggel, M.A. Cazalilla and M. Oshikawa,
						   	{\em Phys. Rev. Lett.} {\bf 107}, 275302 (2011).							
							
\bibitem{landau_1941}		L. D. Landau, 
							{\em Zh. Eksp. Teor. Fiz.} {\bf 11}, 592 (1941).

\bibitem{landau_1947}		L. D. Landau, 
							{\em Zh. Eksp. Teor. Fiz.} {\bf 17}, 91 (1947).
			
\bibitem{taniguchi_2010} 	J. Taniguchi, Y. Aoki and M. Suzuki,
							{\em Phys. Rev. B} {\bf 82}, 104509 (2010).

\bibitem{taniguchi_2011}	J. Taniguchi, R. Fujii and M. Suzuki,
							{\em Phys. Rev. B} {\bf 84}, 134511 (2011).							
							
\bibitem{dalfovo_1995}		F. Dalfovo, A. Lastri, L. Pricoupenko, S. Stringari and J. Treiner,
							{\em Phys. Rev. B} {\bf 52}, 1193 (1995).
					
\bibitem{minoguchi_2011}	T. Minoguchi and D.E. Galli, 
							{\em J. Low Temp. Phys.} {\bf 162}, 160 (2011).
														
\bibitem{minoguchi_2012}	T. Minoguchi, D.E. Galli, M. Rossi and A. Yoshimori, 
 							{\em J. Phys.: Conf. Series} {\bf 400}, 012050 (2012).
							
\bibitem{minoguchi_2013}	T. Minoguchi, M. Nava, F. Tramonto, D.E. Galli, 
							{\em J. Low Temp. Phys.} {\bf 171}, 259 (2013).

\bibitem{bertaina_2016}		G. Bertaina, M. Motta, M. Rossi, E. Vitali and D.E. Galli,
							{\em Phys. Rev. Lett.} {\bf 116}, 135302 (2016).							
							
\bibitem{vitali_2010}		E. Vitali, M. Rossi, L. Reatto and D.E. Galli,
							{\em Phys. Rev. B} {\bf 82}, 174510 (2010).
							
\bibitem{rossi_2012}		M. Rossi, E. Vitali, L. Reatto and D.E. Galli, 
							{\em Phys. Rev. B} {\bf 85}, 014525 (2012).

\bibitem{nava_2013}			M. Nava, D.E. Galli, M.W. Cole and L. Reatto, 
							{\em J. Low Temp. Phys.} {\bf 171}, 699 (2013).

\bibitem{nava_2013b}		M. Nava, D.E. Galli, S. Moroni and E. Vitali, 
							{\em Phys. Rev. B} {\bf 87}, 144506 (2013).

\bibitem{arrigoni_2013}		F. Arrigoni, E.Vitali, D.E.Galli and L. Reatto, 
							{\em Low Temp. Phys./Fizika Nizkikh Temperatur} {\bf 39}, 1021 (2013).
						
\bibitem{rota_2013}			R. Rota, F. Tramonto, D.E. Galli, S. Giorgini, 
							{\em Phys. Rev. B} {\bf 88}, 214505 (2013).

\bibitem{rota_2014}			R. Rota, F. Tramonto, D.E. Galli, and S. Giorgini, 
							{\em J. Phys.: Conference Series} {\bf 529}, 012022 (2014).

\bibitem{molinelli_2016}	S. Molinelli, D.E. Galli, L. Reatto and M. Motta,
							{\em J. Low Temp. Phys.}, {\bf 185} 39, (2016).

\bibitem{sarsa_2000}		A. Sarsa, K.E. Schmidt and W.R. Magro,
							{\em J. Chem. Phys.} {\bf 113}, 1366 (2000).
								
\bibitem{galli_2003}		D.E. Galli and L. Reatto,
							{\em Mol. Phys.} {\bf 101}, 1697 (2003).
								
\bibitem{rossi_2009} 		M. Rossi, M. Nava, L. Reatto and D.E. Galli
							{\em J. Chem. Phys.} {\bf 131}, 154108 (2009).

\bibitem{cherny_2011}		A.Y. Cherny, J.S. Caux, and J. Brand,
							{\em Front. Phys.} {\bf 7}, 54 (2012).

\bibitem{motta_2016}		M. Motta, E. Vitali, M. Rossi, D.E. Galli and G. Bertaina,
							{\em Phys. Rev. A} {\bf 94}, 043627 (2016).

\bibitem{kalos_1974}		M. Kalos, D. Levesque and L. Verlet,
                            {\em Phys. Rev. A} {\bf 9}, 2178 (1974).

\bibitem{astrakharchik_2004}	G.E. Astrakharchik and L.P. Pitaevskii,
								{\em Phys. Rev. A}  {\bf 70}, 013608 (2004).

\bibitem{giamarchi_1988}	T. Giamarchi and H.J. Schulz,
							{\em Phys. Rev. B} {\bf 37}, 325 (1988).

\bibitem{galli_2008}		D.E. Galli and L. Reatto, 
							{\em J. Phys. Soc. Jap.} {\bf 77}, 111010 (2008).

\bibitem{chan_2013}			M.H.W. Chan, R.B. Hallock and L. Reatto,
							{\em J. Low Temp. Phys.} {\bf 172}, 317 (2013).

\bibitem{sugiyama_1992}		G. Sugiyama, C. Bowen and B.J. Alder,
							{\em Phys. Rev. B} {\bf 46}, 13042 (1992).

\bibitem{bowen_1994}		C. Bowen, G.Sugiyama, and B.J. Alder.
							{\em Phys. Rev. B} {\bf 50}, 14838 (1994).

\bibitem{moroni_1995}		S. Moroni, D.M. Ceperley and G. Senatore,
							{\em Phys. Rev. Lett.} {\bf 75}, 689 (1995).

\bibitem{depalo_2003}		S. De Palo, S.Conti and S. Moroni,
							{\em Phys. Rev. B} {\bf 69}, 035109 (2003).	
	
\bibitem{caupin_2008}		F. Caupin, J. Boronat and K.H. Andersen
							{\em J. Low Temp. Phys.} {\bf 152}, 108 (2008).
	
\bibitem{luther_1974}		A. Luther and I. Peschel,
							{\em Phys. Rev. B} {\bf 9}, 2911 (1974).

\bibitem{castro_1994}		A.H. Castro Neto, H.Q. Lin, Y.-H. Chen, and J.M.P. Carmelo,
							{\em Phys. Rev. B} {\bf 50}, 14032 (1994).	

\bibitem{mazzanti_2008}		F. Mazzanti, G.E. Astrakharchik, J. Boronat, and J. Casulleras,
							{\em Phys. Rev. Lett.} {\bf 100}, 20401 (2008).
							
\bibitem{vignale_2005} 		G.F. Giuliani and G. Vignale,
                            {\em Quantum Theory of the Electron Liquid} (Cambridge University Press,2005).
	
\bibitem{motta_2015}		M. Motta, D.E. Galli, S. Moroni and E Vitali,
							{\em J. Chem. Phys.} {\bf 143}, 164108 (2015).
							
\bibitem{ikegami_2007}  	H. Ikegami, Y. Yamato, T. Okuno, J. Taniguchi, N. Wada, S. Inagaki, and Y. Fukushima, 
				{\em Phys. Rev. B} {\bf 76}, 144503 (2007).


\end{thebibliography}
\end{document}